# Nano-Subsidence Assisted Precise Integration of Patterned Two-Dimensional Materials for High-Performance Photodetector Arrays

*Song-Lin Li,[1,2]\* Lei Zhang,[1] Xiaolan Zhong,[1] Marco Gobbi,[1] Simone Bertolazzi,[1] Wei Guo,[3] Bin Wu,[3] Yunqi Liu,[3] Emanuele Orgiu,[1§\*] Paolo Samorì[1\*]*

[1] University of Strasbourg, CNRS, ISIS UMR 7006, 8 allée Gaspard Monge, F-67000 Strasbourg, France

[2] National Laboratory of Solid State Microstructures, School of Electronic Science and Engineering and Collaborative Innovation Center of Advanced Microstructures, Nanjing University, Nanjing 210023, China

[3] Beijing National Laboratory for Molecular Sciences, Institute of Chemistry, Chinese Academy of Science, Beijing 10086, China

[§] *Present address:* Institut national de la recherche scientifique (INRS), EMT Center, 1650 Blvd. Lionel-Boulet, J3X 1S2 Varennes, Canada

**Corresponding Authors** E-mail: sli@nju.edu.cn (S.L.L), emanuele.orgiu@emt.inrs.ca (E.O.), samori@unistra.fr (P.S.)

**ABSTRACT**

The spatially precise integration of arrays of micro-patterned two-dimensional (2D) crystals onto three-dimensionally structured Si/SiO$_2$ substrates represents an attractive strategy towards the low-cost system-on-chip integration of extended functions in silicon microelectronics. However, the reliable integration of the arrays of 2D materials on non-flat surfaces has thus far proved extremely challenging due to their poor adhesion to underlying substrates as ruled by weak van der Waals interactions. Here we report on a novel fabrication method based on nano-subsidence which enables the precise and reliable integration of the micro-patterned 2D materials/silicon photodiode arrays exhibiting high uniformity. Our devices display peak sensitivity as high as 0.35 A/W and external quantum efficiency (EQE) of ~90%, outperforming most commercial photodiodes. The nano-subsidence technique opens a viable path to on-chip integrate 2D crystals onto silicon for beyond-silicon microelectronics.

**MAIN TEXT**

With Moore's law reaching its limit[1], the semiconductor industry is urgently searching for innovative strategies to go beyond standard microelectronics. One of the most recent and intriguing strategies in modern electronics, referred to as system-on-chip (SoC) for more-than-Moore electronics[2-4], aims to integrate various active modules in individual chips wherein different functional materials are combined with standard silicon technology. This hybrid approach enables the introduction of extended functionalities such as data storage, sensing, communication and self-powering to conventional logic modules, therefore expanding the current capabilities of logic microelectronics. In this context, two-dimensional (2D) atomic crystals[5-9], which include a vast library of materials with each one featuring distinctive physical and electronic properties, have emerged as outstanding candidates for integration into silicon to create hybrid devices with unique capabilities[10].

For the purpose of SoC application, different materials are required to be hetero-integrated onto silicon through direct contact or interconnection[4]. Previous reports revealed that 2D crystals could be directly transferred onto three-dimensionally structured silicon for energy harvesting[11-13], photonics[14-16], and electronics[17], providing evidence for their potential as on-chip functional modules. However, in such experiments the cumbersome fabrication process following the 2D crystal transfer needs to be minimized, since the atomically thin 2D crystals are prone to shear from or even to come off the substrate during microfabrication processing such as photoresist deposition and pattern development. These fabrication issues severely hamper the applicability of patterning large 2D sheets into micrometric functional arrays as active SoC components. For these reasons, to date such hybrid devices are mostly restricted to large 2D sheets forming individual active units; the precise and stable integration of arrays of micro-patterned 2D crystals for advanced electronics has remained elusive. Here, we have devised a nano-subsidence integration method which enables spatially precise and high-yield integration of arrays of micro-patterned 2D crystals onto a three-dimensional substrate such as a patterned $Si/SiO_2$ surface. Such method enables the fabrication of high performance arrays of photodetectors. As a proof of concept, 2 × 2 type four-quadrant 2D crystal/silicon diodes used as photodetectors are demonstrated. By engineering the optical antireflection and graphene work function, our photodetectors exhibit remarkably high peak sensitivity up to 0.35 A/W and EQE of ~90% at 480 nm, as well as high spatial uniformity.

Depending on the optical wavelength range, the overall system performances are comparable to or higher than those of commercial silicon diode-based photodetectors.

The conventional method for integrating 2D crystals into silicon-based technology consists in their direct transfer onto three-dimensionally predefined Si/SiO$_2$ substrates[11-15,17,18], as portrayed in Fig. 1a. In this approach, the SiO$_2$ capping above silicon is used as both sacrificing and insulating layers. Selected SiO$_2$ sacrificing areas are pre-etched to open a window, which allows direct contact between the 2D crystals and underlying silicon, serving as active functional region; the rest SiO$_2$ areas around the window work as insulating layers for external wiring, resulting in stepwise substrate surfaces[19]. When 2D crystals are transferred over patterned surfaces, they have to physically adapt to the patterns in relief. Since the 2D crystals cannot uniformly land on the top part(s) of the relief(s) and on the surface during a mechanical transfer process, the poor adhesion between 2D crystals and the substrate leads to the emergence of physical corrugations and bubbles[20,21]. In turn, such weak substrate adhesion causes the sliding off or even removal of 2D crystals in the subsequent lithography processing necessary for patterning the 2D crystals or defining microelectrodes. Moreover, it is extremely challenging to precisely align all the micrometric 2D crystals at the desired positions on pre-patterned substrates, as it relies on manual alignment and operation under optical microscope, where location uncertainty is within several micrometers[17]. Thus, by employing a conventional method, it is almost impossible to integrate arrays of micrometric 2D crystals onto three-dimensional substrates with spatial precision and at high yield.

To address the above-mentioned challenges, we have conceived the nano-subsidence integration method that allows a high-yield hetero-integration (Fig. 1b). The method relies on the metallization before the 2D crystal patterning and SiO$_2$ etching, in order to use the solid electrodes as anchoring bars to prevent the 2D crystals from shearing during the following processing. The alignment among 2D crystals, electrodes, and step edges can then be accurately defined through photolithography, ensuring high spatial precision in the hetero-integration. The third consideration is to etch the SiO$_2$ sacrificing areas in the last fabrication step and the arrays of micro-patterned 2D crystals can gently subside, taking advantage of the flexibility of the ultrathin 2D crystals, and adapt to the stepwise substrates with negligible strain and improved adhesion. As an additional benefit, the formation of hetero-junctions between 2D crystals and silicon takes place in the last step, avoiding exposure of the fresh silicon surface to air thereby avoiding any environmental contamination.

Figure 1c sketches the individual processing steps of the novel integration technique while showing corresponding optical microscopy images. First, the large-area CVD graphene is transferred onto a flat $SiO_2$/Si substrate. Graphene adheres well to the substrate thanks to the large contact area and the superior flatness of the pristine $SiO_2$ surface. Metal electrodes with a given pattern are then deposited onto graphene through a standard photolithographic process followed by thermal evaporation of the metal and a lift-off step. Afterwards, graphene is patterned through photolithography and oxygen plasma etching. Finally, the $SiO_2$ layer in the device active region is wet etched *via* a photoresist mask, so that the graphene layers gently fall down and make contact to the exposed silicon. After thoroughly rinsing in deionized water, the samples are dried out in vacuum to desorb the encapsulated water between graphene and silicon. By means of the capillary action formed during the vacuum evaporation of the encapsulated water, the graphene is in conformal contact with silicon. We found that the crystal/silicon adhesion is readily wetting-enhanced and is much superior to that formed in the simple dry transfer.

As a proof of concept, we apply this method to demonstrate a four-quadrant graphene/silicon photodiode detector, a typical beam position sensing module widely used as collimators and many other adjustment sensors in fiber communication and space guidance[22-24]. The yield of the subsidence integration of graphene to silicon was considerably high: 34 out of 36 quadrants were attained without obvious breaking or folding of graphene, thereby attesting a yield as high as 94 % (supplementary Fig. S2). We note that our approach is extremely versatile and the prototypical 2×2 array fabricated here (Fig. 1c) can be easily upgraded into more complicated arrays/devices.

Alongside the use of a novel integration method, we have paid particular attention towards the engineering energy levels of graphene which is known to be essential for attaining high photosensing performance[11,13]. Figure 2a depicts the energy level alignment and working principle of the graphene/silicon photodiodes. The electron-hole pairs are first generated by light irradiation on silicon; then holes drift into graphene assisted by the built-in electric field at the graphene/silicon interface. Hence, the built-in electric field is the driving force for carrier separation and a high built-in electric field would be favorable for this process. In addition, a large interfacial barrier helps to block the drift of electrons to graphene, reducing the carrier recombination in graphene.

A high-quality CVD graphene[25] with an intrinsic carrier mobility of 2800 cm$^2$V$^{-1}$s$^{-1}$ (supplementary Fig. S3d) was used in this work. To increase the interfacial barrier, we implement p-type doping of graphene *via* spin-coating a thin layer of bis(trifluoromethanesulfonyl)amide (TFSA, [CF$_3$SO$_2$]$_2$NH, inset of Fig. 2b), a strong electron-withdrawing molecule[13,26]. In contrast to the weak doping effect by ambient oxygen and water molecules[27] with a shift of charge neutrality point of 17 V (supplementary Fig. S3a), the TFSA doping determines a larger shift of ~270 V (supplementary Fig. S4c). Such a large shift corresponds to a rather high surface doping concentration of $6.5 \times 10^{13}$ cm$^{-2}$ and a work function increase of 0.75 eV in graphene, as estimated through the equation $\varepsilon_F = \hbar \cdot v_F \cdot \sqrt{\pi n}$, where $\varepsilon_F$, $v_F$, $n$, and $\hbar$ denote Fermi energy, Fermi velocity, carrier concentration, and reduced Planck constant, respectively. The strong doping caused an effective interfacial barrier of 0.79 eV (supplementary Fig. S8d). Accordingly, improved photovoltaic behavior was achieved, as confirmed in Fig. 2b by the current-voltage (*I-V*) characteristics of the pristine and doped devices. TFSA-doped devices exhibited highly enhanced open-circuit voltage ($V_{OC}$), short-circuit current ($I_{SC}$), and fill factor (FF), indicating the critical role of the interfacial barrier on the photovoltaic performance. In Fig. 2c, we analyzed the *I-V* behavior of the doped device under dark condition, which reveals a good ideality factor of 2.02 and a high rectification ratio of 10$^5$ in the bias range of ±1 V, suggesting an excellent photodiode junction quality. The excellent junction quality is also corroborated by the photovoltaic tests. Before antireflection coating, the doped device shows high photovoltaic efficiencies of 10.1% and 9.1% under 520 nm/43.5 mW·cm$^{-2}$ and AM1.5 conditions, respectively (supplementary Fig. S10 a and c), consistent with the literature results.[11-13] The series resistance ($R_s$) value was estimated to be 0.26 Ω/cm$^2$ for the 100×100 μm$^2$ device, which outperforms that of previous reports with device fabricated by direct transfer[11].

We have also measured the response speed of the photodiode under a monochromatic light source. Figure 2d shows a diagram of the piezoelectrically controlled monochromator used in our experiments. The angle of the incoming light from the Xenon lamp is fixed while the angle of reflection beam and the resulting wavelength of the outcoming beam through the slit would change upon piezoelectrical rotation of the optical grating. Since the light power varies with wavelength (the characteristic spectrum of the system is given in supplementary Fig. S11), a modulated photoelectric response would be observed with changing the optical wavelength. Figure 2e displays the modulated photocurrent modes when the incident light wavelength changes from 500 nm to 300, 350, 400 and 450 nm, respectively. At shorter time scales (Fig. 2f) both the rise and decay time are within 500 μs, which represents the fastest signal that can be detected through our experimental setup limited by the millisecond-scale piezoelectric response rate of the grating driver. Hence, the response speed of our devices, of at least 500 μs, certainly represents an underestimate. Generally, the response time would be longer if the density of interfacial states and charge trapping centers were sizeable. The fast response featured by the devices made with our nano-subsidence method suggests that the graphene/silicon interface of the photodiodes is of high quality.

In order to further optimize the performances of our photodiodes, we have employed a surface antireflection capping. This is commonly employed to increase the optical absorption of photodiodes by depositing single or multiple optically transparent antireflection dielectrics[11,28-31]. According to the principle of optical destructive coherence, the reflection rate for a certain light will be minimized when a single antireflective capping layer satisfies the double conditions that $n_{AR} = \sqrt{n_{air} n_{si}}$ and $d_{AR} = \lambda/4n_{AR}$,[28] where $n_{AR}$, $n_{air}$, and $n_{si}$ are the refractive indices of the antireflective capping layer, air and silicon, respectively, and $d_{AR}$ and $\lambda$ are the thickness of the capping layer and the incident light wavelength. Given the atomic thickness of the 2D layers and the low formation energies of lattice defects that make them prone to damages generated by external high-energy atoms,[32-34] capping layers grown *via* aggressive deposition methods (e.g. sputtering) should be avoided. Towards this end, we have used thermally evaporated $MoO_3$ as the capping layer,[30,31] and the damage to 2D layers are expected to be minimized. The values of $n_{MoO3}$ are close to those of $\sqrt{n_{air} n_{si}}$ in most range of visible light (supplementary Fig. S6), suggesting that it is a suitable antireflection material. By theoretical calculations, we also confirmed that the effect of graphene on optical absorption is negligible due to its atomic thickness (supplementary Fig. S7).

Figure 3a sketches the device cross section of pristine versus capped devices and their related optical images. Micro-area reflection measurement revealed that the reflectivity is remarkably reduced from 36% to 8% at λ = 520 nm upon the use of a capping layer of 55-nm MoO$_3$. The reduction of the optical reflection is also corroborated by the difference in brightness of the optical images of the pristine and capped silicon/graphene stacks, where a lower brightness is observed in the latter (around orange dot in the lower panels of Fig. 3a). As a result, the photoresponse of the devices is enhanced. Under 43.5 mW/cm$^2$ illumination at 520 nm, the photocurrent increases from 11.6 to 16.7 mA/cm$^2$ (Fig. 3b). The effect of the MoO$_3$ antireflective capping is further analyzed by correlating micro-area reflection spectra to the external quantum efficiency (EQE) of our devices. Here EQE is estimated by $\text{EQE} = \frac{J_{ph}\hbar C}{eP\lambda}$ where $J_{ph}$ is the photocurrent density, $C$ is the speed of light within vacuum, $e$ is the elementary charge, $P$ is the light power. The EQE data was extracted from the photocurrent spectrum recorded between 320 to 690 nm. The open blue and red circles in Fig. 3c compare the EQE before and after deposition of a 55-nm-MoO$_3$ capping layer. As expected, the capping enhances (or reduces) the EQE around λ = 480 (or λ = 320 nm), in agreement with the interference conditions at the corresponding wavelengths (supplementary Fig. S7c). In Fig. 3c, we also compare the EQE with device absorption rate (1-R, with R the reflectivity) as measured by micro-area reflection. The EQE lines follow closely with the absorption rates, suggesting a near-unity internal quantum efficiency (IQE) since IQE=EQE/(1-R). In the experiments, the peak IQE reaches 90% at 480 nm, approaching the ideal unitary IQE of the photodiodes[28]. The ~10% reduction to the ideal values (supplementary Fig. S7, c and d) stems likely from local variations of refractive index in the MoO$_3$ capping layer caused by the presence of pinholes or oxygen loss. The extremely high IEQ achieved here corroborates again the high integration quality of the arrays *via* our nano-subsidence method.

The photocurrent properties of the capped devices were extensively characterized by varying the optical wavelength and power. Figure 3d plots the photoelectric sensitivity versus the two parameters of light wavelength and power. Despite of strong dependence on light wavelength, the sensitivity is basically power independent within the measurement range of optical power (from 10% to 100%) since the devices show negligible saturation in photoresponse below 100% power (Fig. 3e). To evaluate the potential of the nano-subsided photodiodes as a system-on-chip module, we also compare the EQE of our devices in Fig. 3f and Table 1 with currently commercial photodiodes based on silicon pn junctions (ThorLabs FDS010 and FDS10×10, Hamamatsu S1336BQ). Remarkably, our devices are superior to their commercial counterparts in the visible region from 400 to 700 nm, therefore representing a prototypical high-performance system-on-chip module for the beyond-silicon circuitry.

To assess the photoresponse uniformity of individual array units, we mounted our samples onto a test printed circuit board (PCB) (Fig. 4a). In order to have a reliable (wire) bonding to device electrodes, no $MoO_3$ capping layer is deposited in this test. Figures 4b and 4c show enlarged images on local arrays at different magnification ratios where multiple quadrant photodiodes were prepared. During testing, each unit was exposed to a 532-nm focused laser beam with a power of about 10 W/cm$^2$. Figures 4d-4f show the corresponding photoelectric curves. Almost identical *I-V* curves were recorded for the four units, with $I_{SC}$ =0.56±0.08 μA. A slight degradation of the fill factor was observed in all the four *I-V* curves, which is likely due to the effect of ambient moisture on the hygroscopic dopant TFSA during the measurements.

Finally, we have verified the generalization of the nano-subsidence assisted integration method by replacing graphene with another renowned 2D crystal, *i.e.* molybdenum disulfide ($MoS_2$). Figure 5a shows the optical images of a $MoS_2$/silicon junction before and after $SiO_2$ etching. A mechanically exfoliated 5-layer-thick $MoS_2$ was used in this device. Since $MoS_2$ is normally slightly n-doped, the $MoS_2$/silicon stack can be regarded as an n/n$^{++}$ homo-junction with relatively small barrier heights [28]. Figure 5b shows a corresponding band diagram, where $\Phi_C$ and $\Phi_V$ are the barriers for blocking the reverse motions of electrons and holes at the conduction and valence bands, respectively. Owing to smaller $\Phi_C$ and $\Phi_V$ as compared to the case of doped graphene/silicon (Fig. 2a), a large number of carriers can drift and recombine, resulting in lower $V_{OC}$ and $J_{SC}$ (Fig. 5c) and a reduced fill factor (Fig. 5d). However, in spite of the degraded photoelectric properties with respect to graphene/silicon junctions, the $MoS_2$/silicon junction still exhibited a photoelectric behavior similar to that of small-barrier diodes in which backward current is enhanced and reasonably high other photoelectric parameters. In particular, a large rectification ratio of 10$^5$ within ±1 V (Fig. 5c), notable photoelectric behavior (Figs. 5c and 5d), reasonably high peak EQE of ~25% (Fig. 5g), and fast photoresponse of 1 ms (Fig. 5h) are measured.

In conclusion, we have developed a novel method enabling the integration of atomically thin 2D crystals for the system-on-chip electronics, through the precise location of ultrathin crystals and on-chip integration of arrays of micro-patterned 2D crystals with a minimized residual strain of the 2D crystals on non-flat substrates. Noteworthy, this integration method combines several advantages. First, the substrate surface was kept flat in all photolithography steps (*i.e.*, metallization, patterning 2D crystals, and defining mask for $SiO_2$ etching), which facilitates the microfabrication processing such as resist coating. Second, high-yield integration of 2D crystals into the stepwise substrate was realized *via* controlled gentle subsidence assisted by capillary forces during vacuum dry, which also ensures a precise positioning and alignment of the micrometric 2D crystals onto the target electrodes. Third, the exposure of bare silicon surface to air is minimized, resulting in a high quality of the 2D material/Si interface which, as a result, improves the junction performances. Taking graphene and $MoS_2$ as model systems, we demonstrated the general applicability of such unique integration method to integrate all kinds of 2D crystals onto silicon as photodiodes. The photodetector performances surpass those of commercial photodiodes after appropriate device optimization. Not limited to the photoelectric function and materials demonstrated above, the concept of subsidence integration *via* an underlying sacrificial layer could also be extended to wider applications, such as 3D interconnection, optical waveguides, and microfluidic channels, and hence it holds great potential for realizing more versatile modules for the beyond-silicon more-than-Moore microelectronics.

**Acknowledgments**

We acknowledge funding from the European Commission through the Graphene Flagship (GA-696656), the FET project UPGRADE (GA-309056) and Marie-Curie IEF MULTI2DSWITCH (GA-700802), the M-ERA.NET project MODIGLIANI, the Agence Nationale de la Recherche through the Labex projects CSC (ANR-10-LABX-0026 CSC) and Nanostructures in Interaction with their Environment (ANR-11-LABX-0058 NIE) within the Investissement d'Avenir program (ANR-10-120 IDEX-0002-02), and the International Center for Frontier Research in Chemistry (icFRC). This project is also partially supported by the National Key R&D Program of China (2017YFA0206304) and the National Natural Science Foundation of China (61674080).

**Author contributions**

S.L.L., E.O. and P.S. conceived the experiment and designed the study. S.L.L. performed the experiments and developed the fabrication method. W.G., B.W, and Y.L. synthesized and provided the CVD graphene sample. S.L.L., E.O. and P.S. co-wrote the paper. All authors discussed the results and contributed to the interpretation of data, as well as contributing to editing the manuscript.

**Additional information**

Supplementary information is available in the online version of the paper. Reprints and permissions information is available online at www.nature.com/reprints. Correspondence and requests for materials should be addressed to S.L.L., E.O. and P.S.

**Competing financial interests**

The authors declare no competing financial interests.

**Table 1 Comparison of EQE performance of nano-subsidence fabricated graphene/silicon photodiode with typical commercial silicon photodiodes at different wavelength values. The values in parentheses show the EQE difference of the commercial devices as compared to our graphene/silicon device.**

| Photodiode | EQE @ 350 nm | EQE @ 450 nm | EQE @ 500 nm | EQE @ 550 nm | EQE @ 650 nm |
|---|---|---|---|---|---|
| This work | 27.2% | 84.2% | 90.3% | 89.0% | 86.2% |
| ThorLabs FDS10×10 (UV enhanced) | 58.9% (+117%) | 68.6% (-19%) | 73.6% (-18%) | 76.5% (-14%) | 79.0% (-18%) |
| Hamamatsu S1336BQ (UV enhanced) | 56.1% (+106%) | 60.8% (-28%) | 63.7% (-29%) | 65.7% (-26%) | 67.5% (-22%) |
| ThorLabs FDS010 | 19.0% (-30%) | 40.8% (-52%) | 54.2% (-40%) | 64.5% (-28%) | 76.8% (-11%) |

## Methods

**Transfer of CVD graphene on flat SiO$_2$/Si substrates.** Large-area high-quality monolayer graphene was grown on 25 μm thick copper foils by CVD [25]. A 20 mg/mL PMMA/chlorobenzene solution was spin-coated on the graphene/copper foils at 3000 rpm for 30 s, which was then heated dry on a hot plate at 180 °C for 1 minute. A PDMS scaffold with a hole of ~10 mm in diameter was gently pressed down onto the PMMA/graphene/copper stacks, with the PDMS scaffold attaching to the stacks. Afterward, the whole stack was placed floating on an ammonium persulfate (0.1 M) solution with the copper face downwards to etch the copper foil. After removing the copper foils, the PDMS/PMMA/graphene stack was rinsed in distilled water for several times and was finally scooped out by a flat SiO$_2$/Si substrate. The silicon wafers (from IPMS Fraunhofer Institute, Dresden) were capped with a 90-nm-thick thermally grown SiO$_2$ dielectric layer and were n-doped to a high level of $\sim 3 \times 10^{17}$ cm$^{-3}$.

**Device fabrication.**

Optical lithography was performed through a direct laser writing system (LW405B, Microtech Inc.). A thin positive photoresist AZ1505 was used as a mask for graphene patterning and metallization. The exposure resolution is about 1 μm. AZ 726 metal-ion free developer and dimethyl sulfoxide were used for resist development and lift-off, respectively. To increase the adhesion ability of resist and development, the surface of silicon wafers was modified by thermally evaporated hexamethyldisilazane molecules before applying the resist. The SiO$_2$ dielectric layers were etched by standard buffered HF etchant (NH$_4$F : HF = 6 : 1). The electrodes were realized by thermal evaporation of 1 nm of chromium and 50 nm of gold. The use of chromium adhesion layer is necessary in order to prevent the unwanted lateral etching the SiO$_2$ underneath electrodes (See also supplementary Fig. S1).

**Characterization and measurements.**

All optical images of graphene and devices were taken with an Olympus BX53M microscope. The monochromatic photoelectric characterization was performed with a monochromatic source (model Polychrome V, Till Photonic Inc.) and a Cascade

EPS150TRIAX Probe Station inside a nitrogen-filled glovebox. The optical spectrum and its power were calibrated by a PM100A Power Meter (Thorlab) and the related data is shown in supplementary Fig. S11. For the characterization of the integration uniformity of the four-quadrant photodetectors, the device was mounted on a home-made PCB chip carrier which was placed under a focused 532-nm laser beam (~10 W/cm$^2$) from a Renishaw Raman spectroscopy in ambient. A dual-channel sourcemeter Keithley 2636A was used for the electrical characterization.

Published at *ACS Nano* 13, 2654 (2019), doi: 10.1021/acsnano.9b00889

Figure TOC

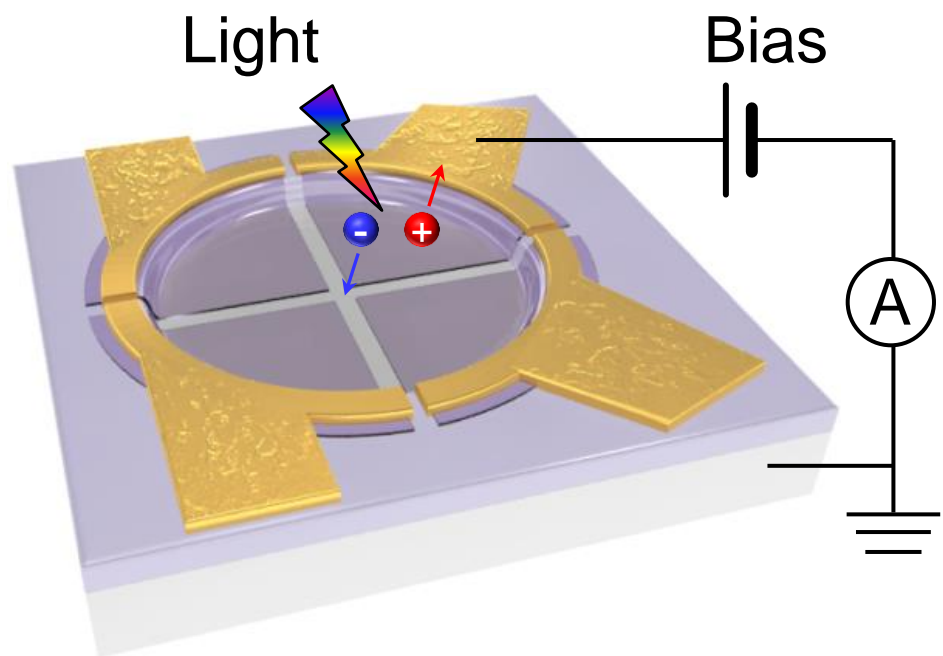

For Table of Contents Only

# Figure 1

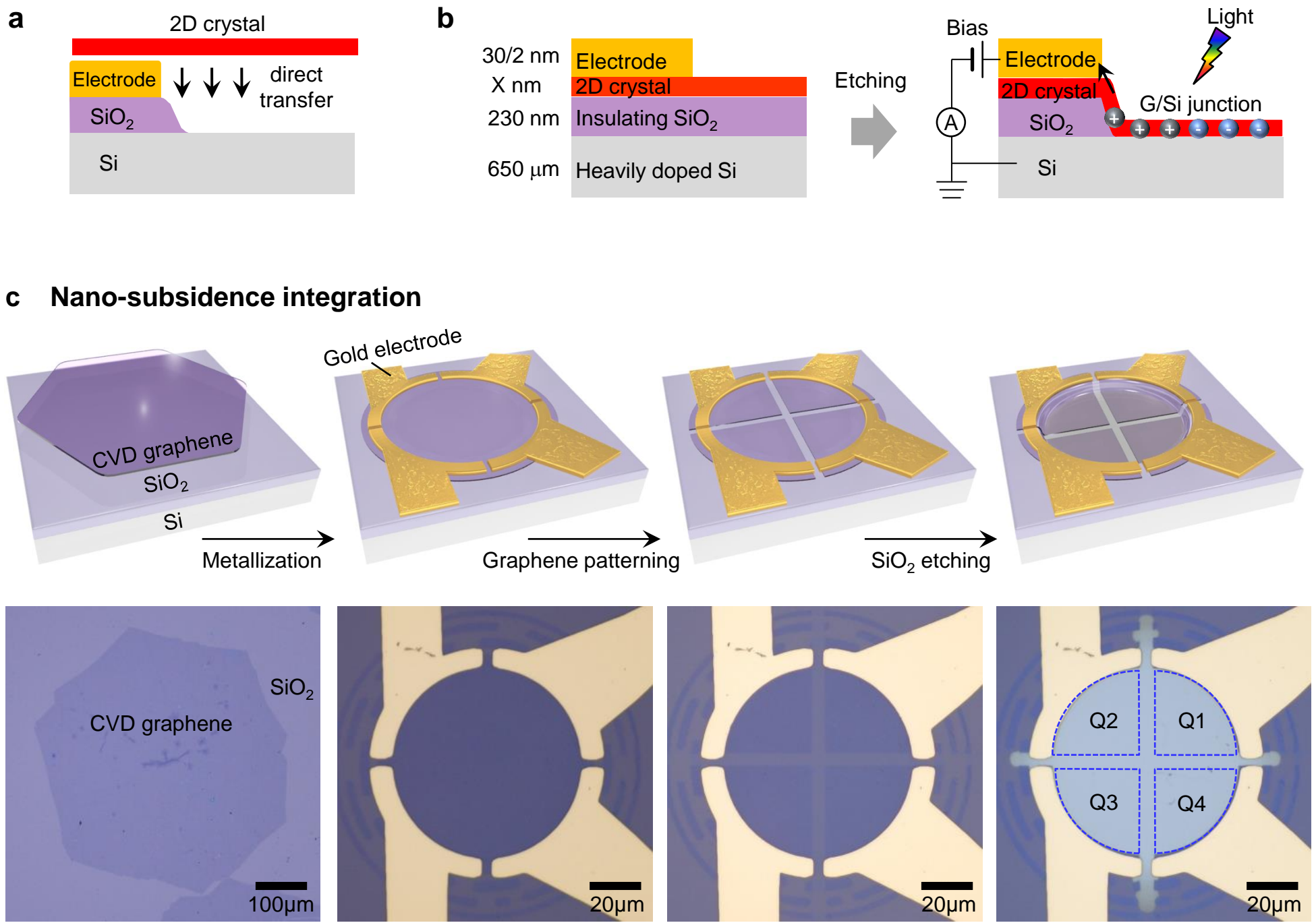

**Figure 1 Schematic diagrams of hybrid integration by conventional direct transfer and our nano-subsidence techniques. a**, Cross-sectional diagram for the conventional direct transfer integration, which features several risks of tearing out and sliding off when the sizes of 2D crystals are small, due to the weak stickability of 2D crystals and the presence of stepwise substrate structure. **b**, The concept of the improved hybrid integration by nano-subsidence in which the 2D crystals are fixed by using the metallic electrodes as anchoring bars and selectively etching out the sacrificial $SiO_2$ layer in the last. The hybrid integration is completed after the gentle subsidence of 2D crystals. **c**, The processing flow and corresponding images for each critical integration step for the subsidence integration. Its processing sequence is renewed to transfer-metallization-patterning-etching to ensure the precise location of 2D crystals.

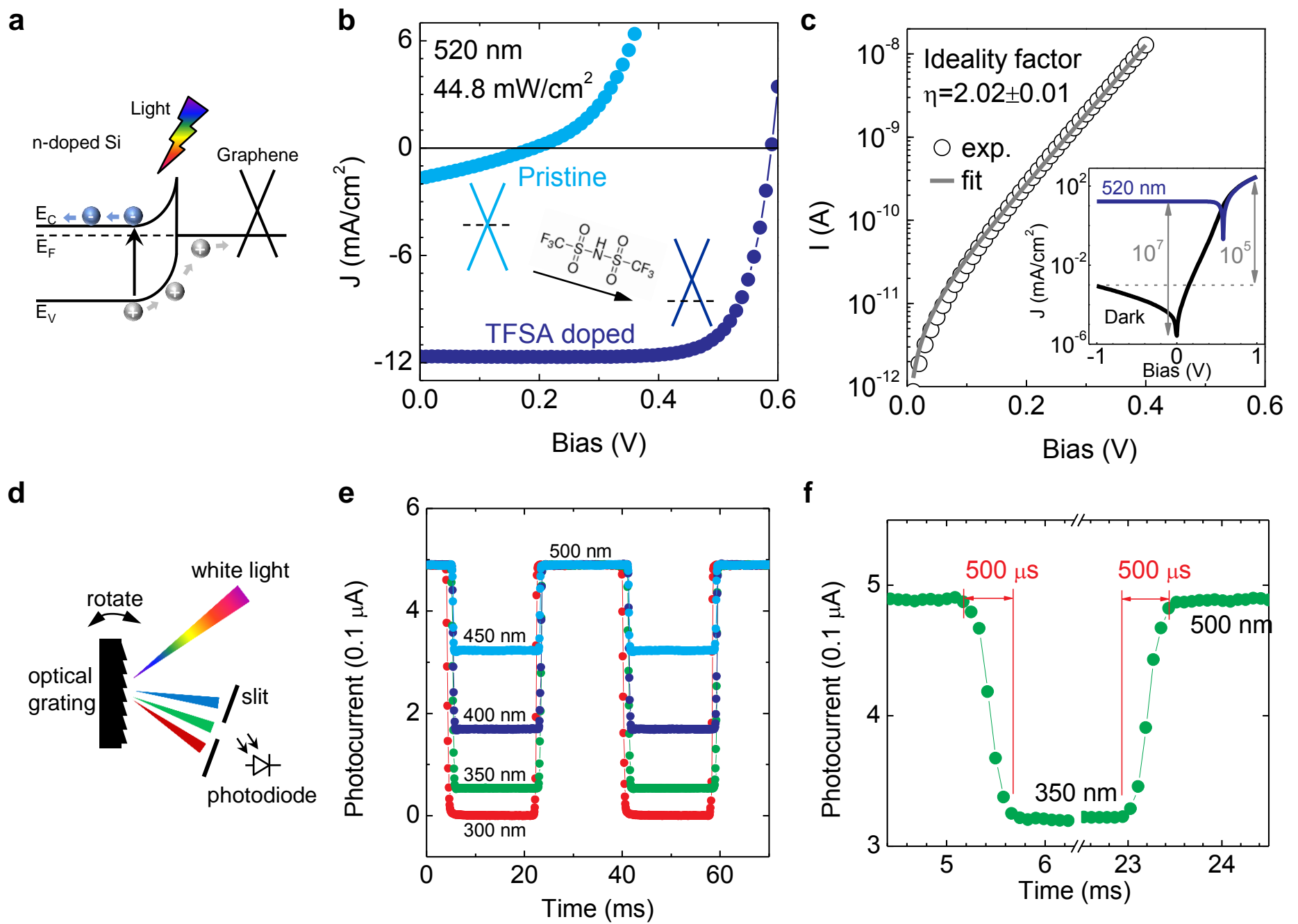

**Figure 2 Photoelectric properties of hybrid graphene/silicon diodes prepared by the nano-subsidence integration. a**, Diagram of the energy level alignment and operation principle of the graphene/silicon photodiodes. **b**, Comparative photoelectric behavior of the photodiodes before and after engineering band alignment *via* TFSA doping. Inset: The molecular structure of TFSA and the change of Fermi level of graphene before and after TFSA doping. **c**, Fitting of *I-V* curve under dark condition to extract the ideality factor of the diodes. Inset: Semi-logarithmic plot of the I-*V* curves under both dark and light conditions. High rectification and zero-bias signal-noise ratios of $10^5$ and $10^7$ are observed. **d**, Principle of wavelength scan for characterizing photoresponse time, where the piezoelectric response time is below the order of ms that defines the lower limit of our samples. **e**, Modulation of photocurrent by varying the excitation wavelength. **f**, Enlarged figure to analyze the rise and decay times of our photodiodes which are estimated to be better than 500 μs.

Figure 3

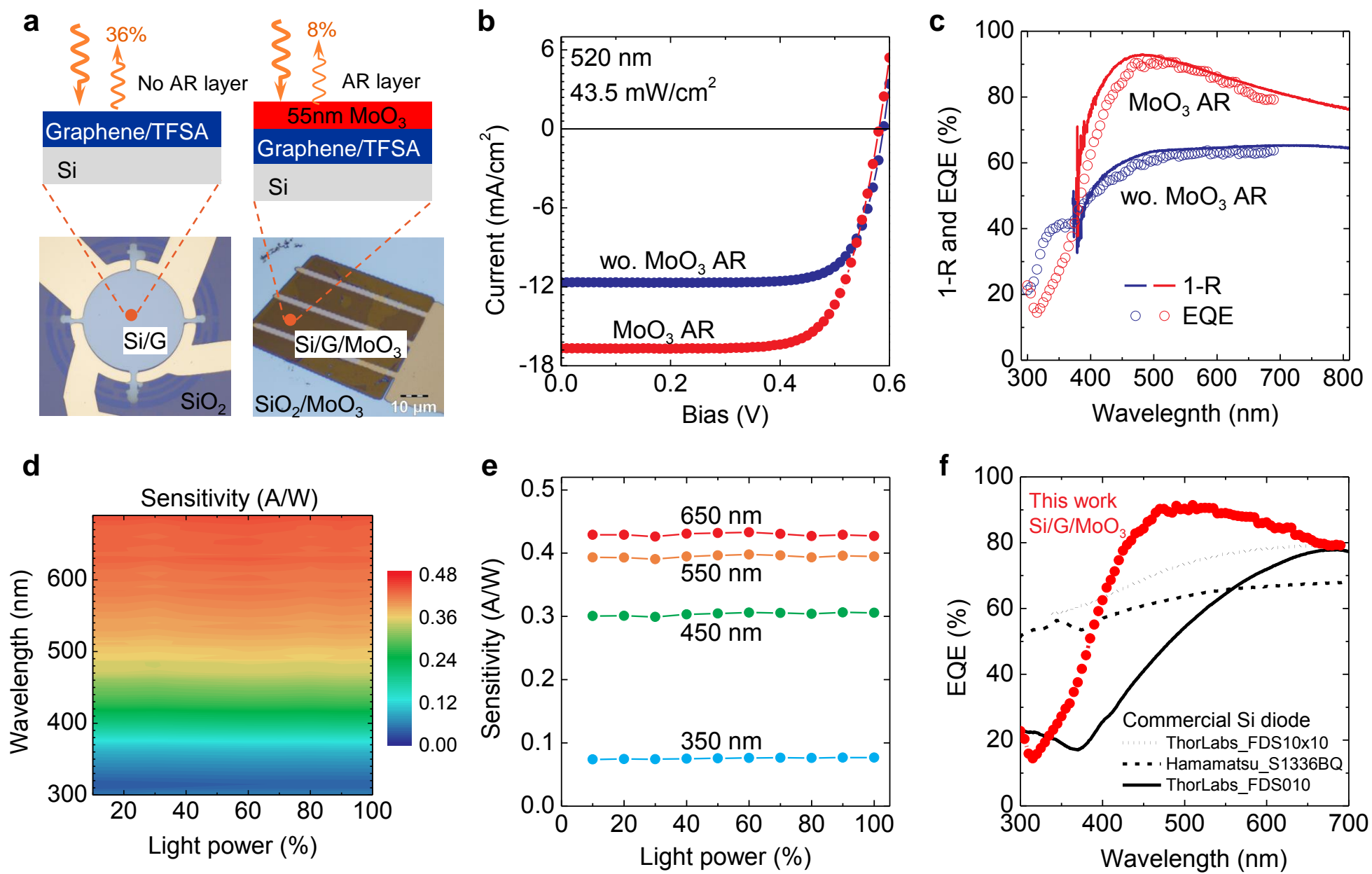

**Figure 3 Photoelectric properties of graphene/silicon diodes after capping antireflective MoO₃.** **a**, Cross-sectional diagrams and optical images of the pristine and MoO₃ capped devices. **b**, Comparative *I-V* behavior of the photodiodes before (blue) and after (red) capping 55 nm MoO₃ antireflective layers. The photocurrent increases from 12 to 17 mA/cm². **c**, Comparison of absorption rate (1-R, lines) and external quantum efficiency (EQE, open dots) before (blue) and after (red) MoO₃ capping. **d**, Contour plot of photoelectric sensitivity *versus* wavelength and light power. **e**, Sensitivity as a function of light power at different wavelength values from 350 to 650 nm. The devices show negligible saturation in photoresponse within experimental power range of ~50 mW/cm². **f**, Comparison of EQE with three typical commercial silicon photodiodes. Our devices (red dots) rivals the counterparts in the visible regime from 400 to 700 nm after the MoO₃ antireflective capping.

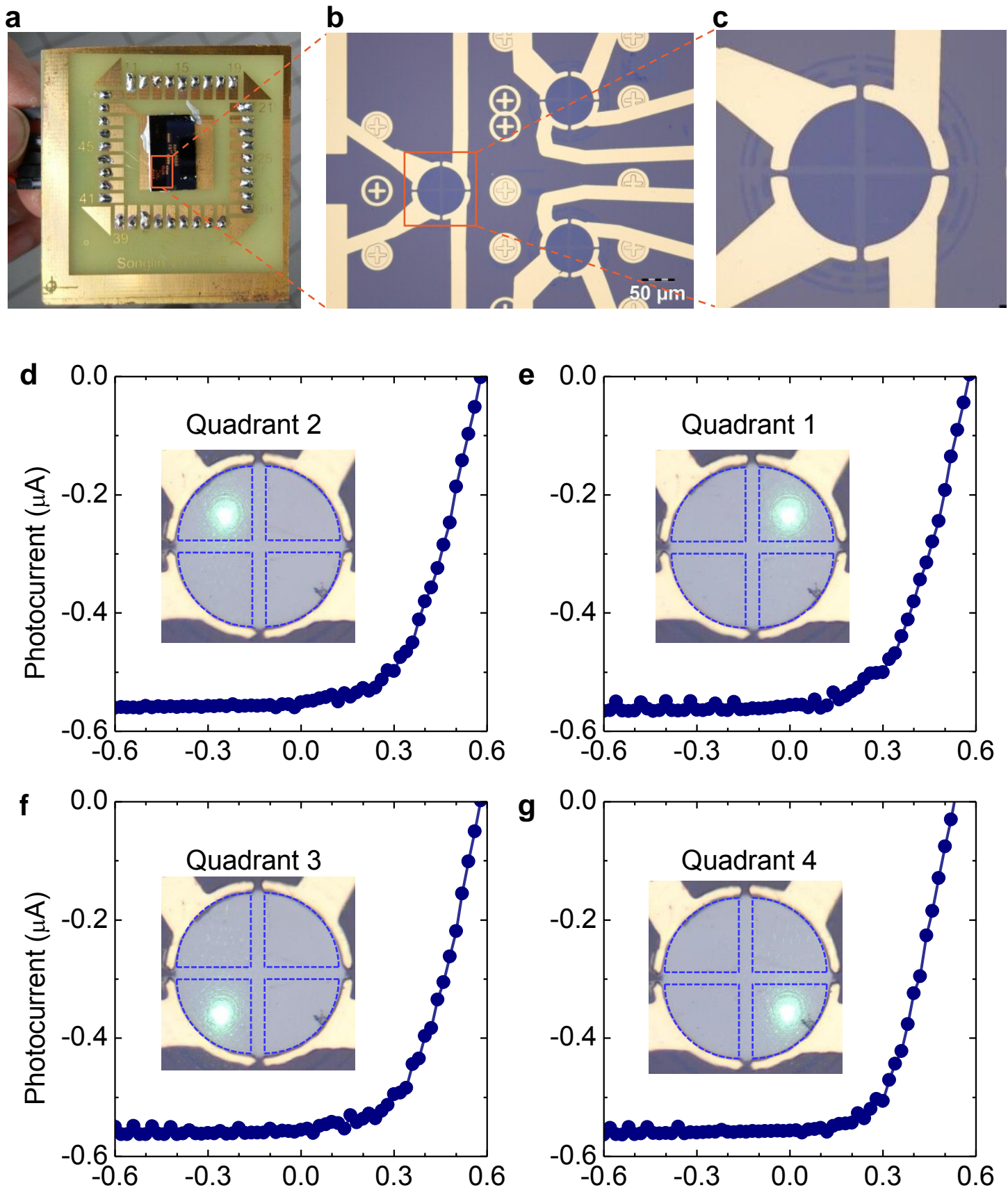

**Figure 4 Test of the uniformity of array units. a**, Optical image of a testing module with samples mounted onto a home-made printed circuit board. **b**, Enlarged image for a local area with three quadrant arrays. **c**, Further enlarged image for an individual 2×2 quadrant array. **d-g**, One-by-one test of the photoelectric behavior for the four array units (*i.e.*, from quadrant 1 to 4). Inset images show the illumination locations of the focused excitation laser.

# Figure 5

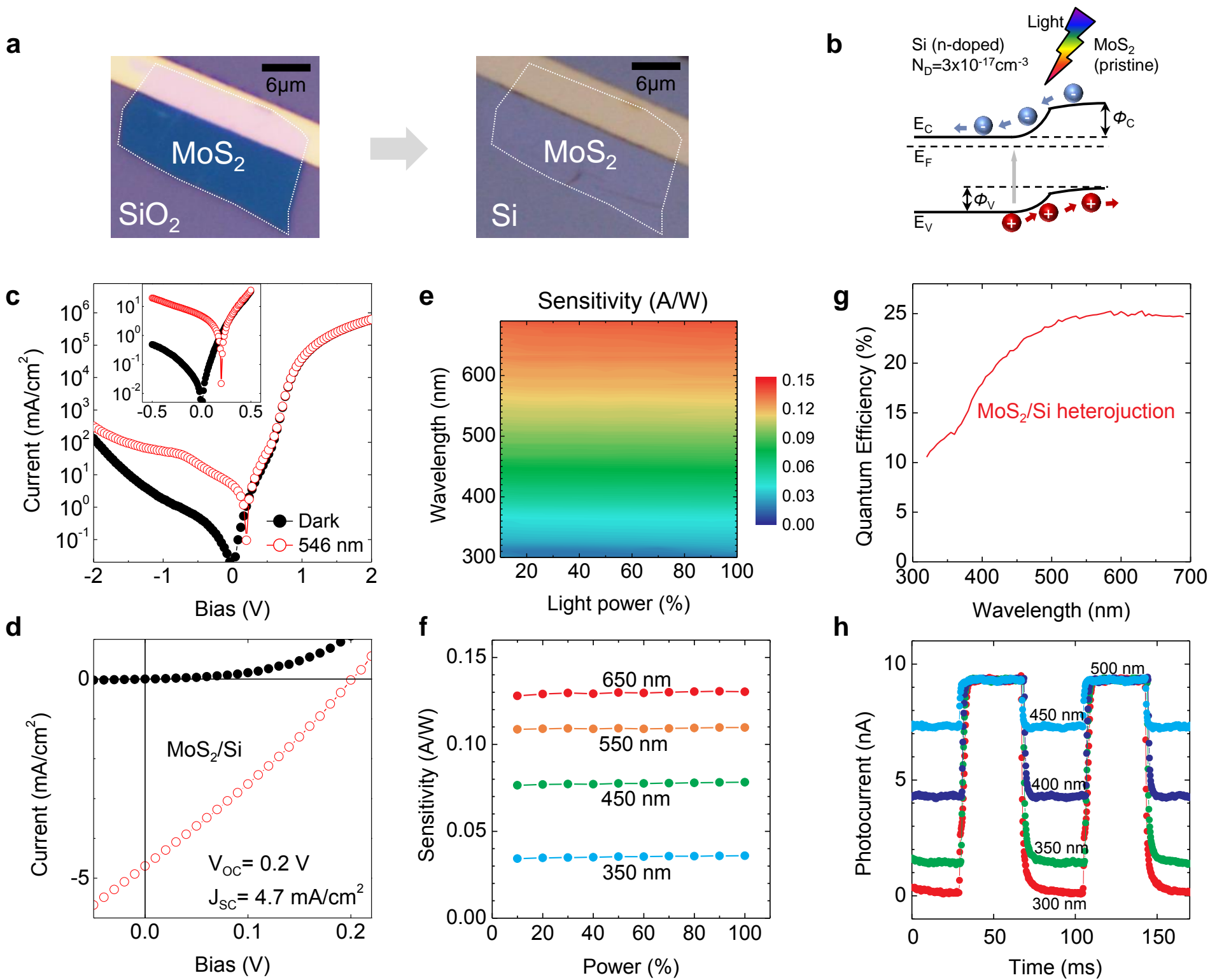

**Figure 5 Test of the feasibility of the subsidence integration technique to other 2D crystals.**
**a**, Optical images for a typical MoS$_2$/silicon diode before and after SiO$_2$ etching. **b**, Diagram of the energy level alignment of the MoS$_2$/silicon diode, which is actually an n/n$^{++}$ heterojunction with small barrier heights ($\Phi_C$ and $\Phi_V$). **c**, Semi-logarithmic plot of the *I-V* curves under both dark (black dots) and light (red circles) conditions. **d**, Corresponding linear plot of the *I-V* curve. **e**, Contour plot of photoelectric sensitivity *versus* wavelength and light power. **f**, Sensitivity as a function of light power at different wavelength values from 350 to 650 nm. The devices also show negligible saturation in photoresponse within experimental power range. **g**, Estimated EQE for different wavelengths from 320 to 700 nm. **h**, Modulation of photocurrent by varying the excitation wavelength.

*Supporting Information*

**Nano-Subsidence Assisted Precise Integration of Patterned Two-Dimensional Materials for High-Performance Photodetector Arrays**

*Song-Lin Li,*[*,1,2,5,6] *Lei Zhang,*[1] *Xiaolan Zhong,*[1] *Marco Gobbi,*[1] *Simone Bertolazzi,*[1] *Wei Guo,*[3] *Bin Wu,*[3] *Yunqi Liu,*[3] *Ning Xu,*[2] *Weiyu Niu,*[4] *Yufeng Hao,*[4,5,6] *Emanuele Orgiu,*[*,1,§] *Paolo Samorì*[*,1]

[1]University of Strasbourg, CNRS, ISIS UMR 7006, 8 allée Gaspard Monge, F-67000 Strasbourg, France; [2]School of Electronic Science and Engineering, Nanjing University, Nanjing 210023, China; [3]Beijing National Laboratory for Molecular Sciences, Institute of Chemistry, Chinese Academy of Science, Beijing 10086, China; [4]College of Engineering and Applied Sciences, Nanjing University, Nanjing 210023, China; [5]National Laboratory of Solid State Microstructures, Nanjing University, Nanjing 210093, China; [6]Collaborative Innovation Center of Advanced Microstructures, Nanjing University, Nanjing 210093, China; [§]*Present address:* Institut national de la recherche scientifique (INRS), EMT Center, 1650 Blvd. Lionel-Boulet, J3X 1S2 Varennes, Canada

**Address correspondence to** sli@nju.edu.cn (S.L.L), emanuele.orgiu@emt.inrs.ca (E.O.), samori@unistra.fr (P.S.)

**Table of Content**

# 1. Etching rates of SiO₂ at different interfaces

For making 2D flake/Si junction with nano-subsidence method, there are three types of SiO$_2$ interfaces deserved to consider during the SiO$_2$ etching: 2D flake/SiO$_2$, electrode/SiO$_2$, and resist mask/SiO$_2$. The first one is the desired interface for the penetration of SiO$_2$ etchant (BOE solution) to remove the SiO$_2$ underneath the 2D flake, while the latter two are not desired. Hence, the lateral etching rates for the latter two should be minimized. We found that the factor to determine the lateral etching rate of the material/SiO$_2$ stacks is the chemical bonds formed between the top capping layer and the bottom SiO$_2$. In general, the rate is small when there are chemical bonds (Cr/SiO$_2$, resist mask/primed SiO$_2$) while it becomes large when no bonds are present (graphene/SiO$_2$, Au/SiO$_2$, resist mask/unprimed SiO$_2$). Since there are no chemical bonds formed between the 2D flake/SiO$_2$ stacks, the lateral etching rate is rather large. For the electrode/SiO$_2$ stacks, an adhesion metal layer such as chromium is very necessary to minimize the detrimental lateral etching of the SiO$_2$ under electrodes.

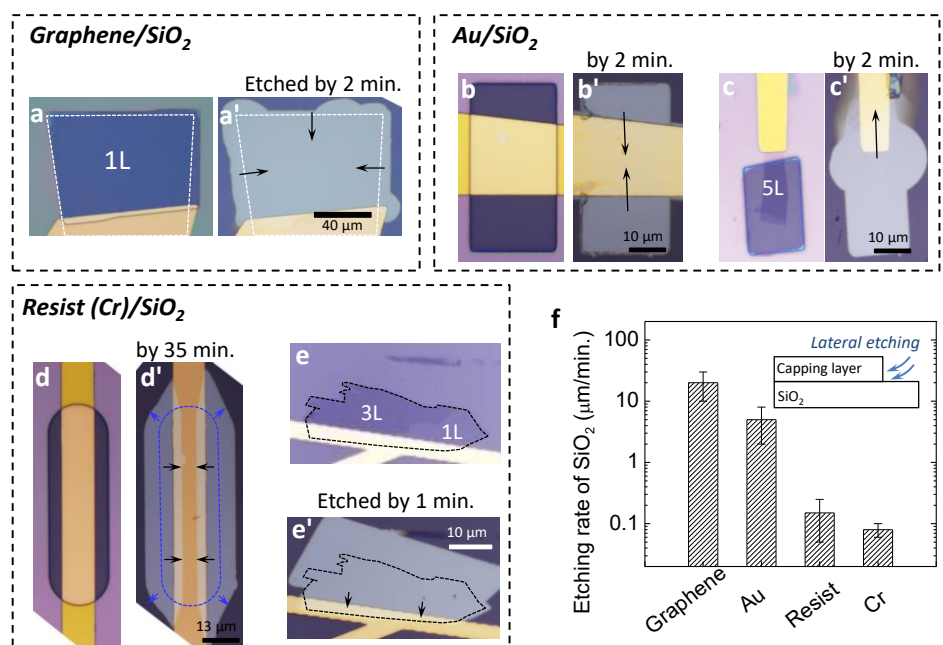

**Figure S1 Etching rates of SiO₂ underneath different capping layers. a-e**, Optical images of devices just before SiO$_2$ etching; **a'-e'**, Corresponding images for devices after etching. The dashed black lines and black arrows indicate the area of graphene and etching directions, respectively. The dashed blue line and blue arrows in panel **d'** denote the patterned resist windows for etching and the direction of etching progression of the resist mask, respectively. **a**, graphene capped SiO$_2$; **b** and **c**, Au capped SiO$_2$; **d** and **e**, resist and Cr capped SiO$_2$. **f**, Estimated etching rates for SiO$_2$ capped by four types of materials (graphene, Au, mask resist and Cr). Inset: Diagram of lateral etching of SiO$_2$ underneath a capping layer.

Figure S1 shows the results of a comparative experiment where the etching rates of SiO$_2$ at different interfaces (*i.e.*, graphene/SiO$_2$, Au/SiO$_2$, resist mask/SiO$_2$ and Cr/SiO$_2$) are measured. The panels a-e show images of the samples before etching (etching windows to be exposed are visible); the panels a'-e' show the images after etching. The dashed lines and black arrows indicate the area of graphene and etching directions, respectively.

It is well known that there is only simple physical contact (van der Waals interaction) between either graphene or pure gold electrode and SiO$_2$ (*i.e.*, graphene/SiO$_2$ and Au/SiO$_2$), where no chemical bonds form, and thus the lateral etching rate of SiO$_2$ is high. Within 2 minutes the etching length could reach as much as 40 and 15 μm at the interfaces of graphene/SiO$_2$ and Au/SiO$_2$, respectively. On the contrary, there are strong chemical bonds between Cr and SiO$_2$ or between resist mask and HMDS primed SiO$_2$ surface; hence the etching rates for them are as low as ~0.1 and ~0.15 μm/min, respectively. By constant soaking in HF solution of 35 minutes, the etching lengths of SiO$_2$ are only ~3 and 5.2 μm in cases of Cr and resist mask capping (panels d and d'). Within the normal etching time of about 1.5 min, the unwanted etching lengths are 200-300 nm for the graphene underneath the Au/Cr electrode and resist mask (panels e and e'); the detrimental etching of SiO$_2$ is negligible.

There is also potential risk of formation of low resistive pathways caused by the collapse of the electrode/2D crystal bilayer. However, we found that there is some tolerance in experiment. First, the work function of the Cr/Au electrode (1 nm Cr, 50 nm of Au) is relatively high, mainly following the thick Au layer, and it would form a Schottky junction for the trilayer structure electrode/graphene/n doped silicon. Hence, no obvious low resistive pathway/connection appears even in the event of a small-area bilayer collapse. Second, as an electronic effect of the collapse, the contacting area results in the increase of unfavorable shunt resistance in the circuit loop of photodiodes, which may deform the I-V curve or short-circuit the devices, depending on the magnitude of the shunt resistance. In our experiment, we observed that most devices remain electronically functional with 2-3 μm over etching. The stopping point of etching progression is just the boundary of the 2D crystals, outside which the Cr adhesion layer is strongly bonded to underlying SiO$_2$ and results in negligible lateral etching rate of this area, as shown by the dashed line in Fig S1 e and e'.

Panel f lists the lateral etching rates for the three interfaces. It is shown that the etching rate can vary by 2 orders of magnitude depending on the interface condition *i.e.*, whether chemical bonds can be formed at the interfaces. For instance, depositing

Au or graphene onto SiO$_2$ will not create chemical bonds between Au (graphene) and SiO$_2$, which results in a high lateral etching rate. Instead, depositing Cr onto SiO$_2$ creates an additional Cr-O bonds (in presence as chromium oxide) between Cr and SiO$_2$, which leads to a low lateral etching rate. Since the processing step of SiO$_2$ etching is placed in the last in our nano-subsidence technique, a precise control of the etching rate of SiO$_2$ is critical for the overall fabrication resolution. We verified that the addition of the adhesion layer (such as Cr, Ti) underneath the metallic electrode, which is translated as an adhesion layer between the electrode and SiO$_2$ during etching (Figure 1b in the main text), can avoid the unfavorable etching of SiO$_2$ and hence ensure our fabrication resolution.

## 2. Nano-subsidence integration

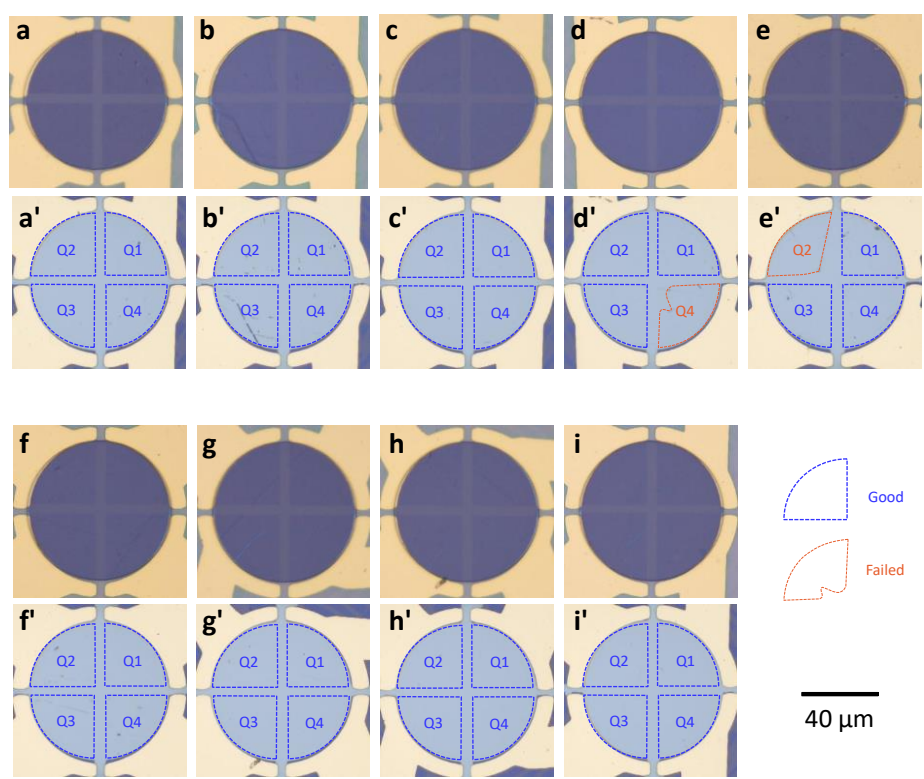

**Figure S2 Yield of device fabricated by the nano-subsidence technique. a-i**, Optical images of the four-quadrant photodetector before nano-subsidence; **a'-i'**, Corresponding images for devices after applying the nano-subsidence. The blue and orange dashed lines represent the good and failed quadrants, respectively. Among the 36 quadrants, 34 of them are good.

Although slightly fierce operations (such as water scouring to graphene during rinse after HF etching) are employed in sample fabrication, most graphene pieces stick well to the underlying Si layers. Figure S2 show the optical images for the 9 arrays of the four-quadrant devices before and after the subsidence processing. Here, we take one quadrant as an individual sample and the quadrants with visible graphene folding, scrolling or rupture are counted as the failure. We estimated the sample yield as high as 94.4% (34 out of 36 quadrants), which indicates the high reliability of our nano-subsidence technique.

By using conventional photolithography, we also fabricated large scale arrays with the nano-subsidence and conventional integration methods, respectively, to enable direct comparison of the two methods. The main results are listed in Table S1 with the superior figures of merit marked in bold. The reported values provide unambiguous evidence that our nano-subsidence integration method outperforms in parameters such as sample yield, probability of strain accumulation, surface roughness of graphene/silicon, and time of air exposure of bare silicon, although both integration methods result in comparable peak photoresponsivity.

We have recorded AFM images of the surface following nano-subsidence and conventional integration techniques. The results are displayed in Figure S3 below. We used the root mean square (RMS) roughness in order to gain a quantitative insight into the graphene morphology, and to evaluate the cleanness of the graphene/silicon interfaces. The left panel, displaying devices obtained using our nano-subsidence integration method, reveals a cleaner graphene interface with lower RMS roughness of ~0.52 nm, compared to the right panel which displays interfaces obtained using conventional method, the latter exhibiting a RMS roughness of ~0.84 nm.

**Table S1** Comparison of different parameters of the two integration methods. Better values are indicated in bold.

|  | *Nano-subsidence method* | *Conventional method* |
|---|---|---|
| Sample yield | **~96%** | ~88% |
| Probability of strain accumulation | **<1%** | >8% |
| Surface roughness of graphene/silicon | **0.52 nm** | 0.84 nm |
| Time of air exposure of bare silicon | **< 60 seconds** | >3 hours |
| Peak photoresponsivity | **0.35 A/W** | **0.37 A/W** |

First, the 8% higher yield ratio (96% *vs.* 88%) is attributed to the effect of electrode anchoring, which stabilizes one of the three edges of the slidable graphene flakes during subsequent microfabricaton processing.

Second, the greatly reduced probability of strain accumulation (from 8% to 1%) can be ascribed to the different integration process. In the conventional way, entire pieces of graphene sheets are transferred and closely attached to the whole substrates; no more chance for strain release after heat dry. Instead, in our new technique, the strain accumulated in the first transfer process can be well released in the process of nano-subsidence when $SiO_2$ is gradually etching way. As mentioned above, except for the anchored edge (Figure 1c, bottom panels), there are two edges of the graphene flakes floated in the etchant and the strain can be well released during the subsidence to bare silicon.

Third, the difference in interface cleanness (surface roughness, Figure S3) of graphene/silicon (0.52 *vs.* 0.84 nm) likely arises from the specific device microstructure as depicted in the cartoon. The devices from nano-subsidence generally feature a less rough underlying support for graphene, which facilitates the expulsion of liquids and soluble impurities during heat/vacuum dry. Instead, the devices fabricated following the conventional method are predefined with silicon cavities, which therefore tend to trap soluble impurities and to contaminate the graphene/silicon interface after heat dry. Trapping of impurities led to an increase of surface roughness.

Other bonus from the nano-subsidence technique includes minimized time of air exposure of bare silicon and preserved fresh silicon interface, which offer great flexibility in further interface engineering between silicon and 2D crystals.

We also need to clarify the device performance from the two integration techniques, which are comparable, since the peak photoresponsivity reaches 0.35-0.37 A/W when other device parameters are fully optimized. This indicates that the device performance has little to do with the integration methods.

Finally, it is worth noting that the yield ratio of the $MoS_2$ junctions is slightly higher than graphene junctions, reaching nearly 97%, because the $MoS_2$ flakes adopted are generally thicker and more rigid than graphene, leaving them little chances to rupture.

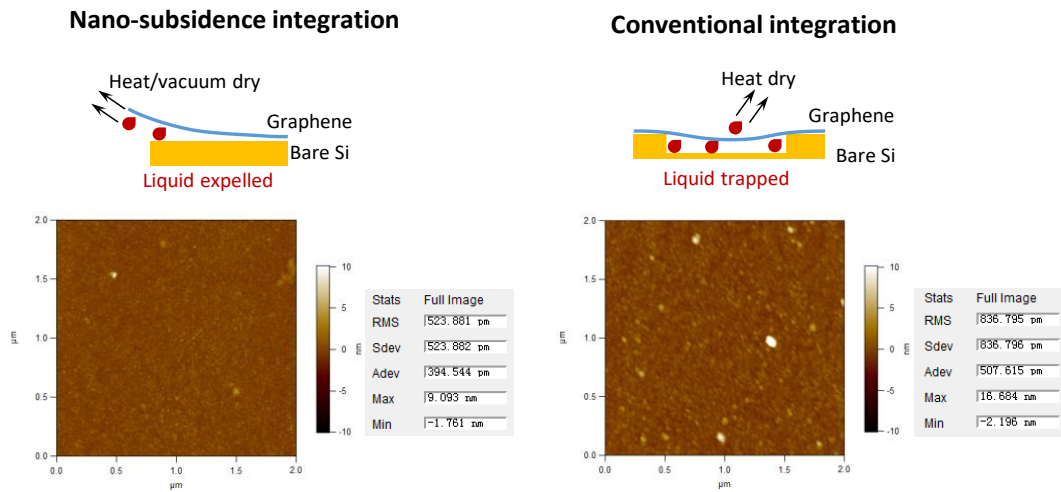

**Figure S3** Schematics of formation of interfacial residues between graphene and bare silicon and the corresponding surface roughness of graphene after $SiO_2$ removal as seen with AFM. Left and right panels are for the nano-subsidence and conventional integration techniques, respectively.

Since thick 2D materials would become rigid and unbendable, the conformal step coverage of thick 2D materials on patterned Si substrates would be difficult and there are certain thickness limits for using the nano-subsidence integration. From our data, we can conclude that the lower thickness limit for subsidence integration is 4 layers for graphene and 5 layers for $MoS_2$ (Figure 5a), respectively. Another factor alleviating the requirement on thickness limit is the gentle slope of ~16° formed at the step, as confirmed by the AFM profile and depicted by the cartoon (Figure S4, right panel). The gentle, rather than sharp, slope normally facilitates the conformal step coverage of 2D films on the patterned steps.

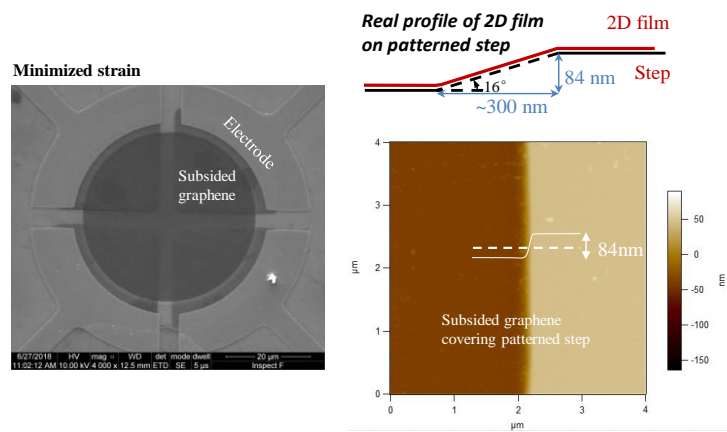

**Figure S4** Left: SEM images for junctions with well-subsided graphene. Right: AFM image near a patterned edge of the graphene junction. In both images, no clear traces of strain induced ripples are observed.

## 3. Intrinsic mobility and contact resistivity of pristine CVD graphene

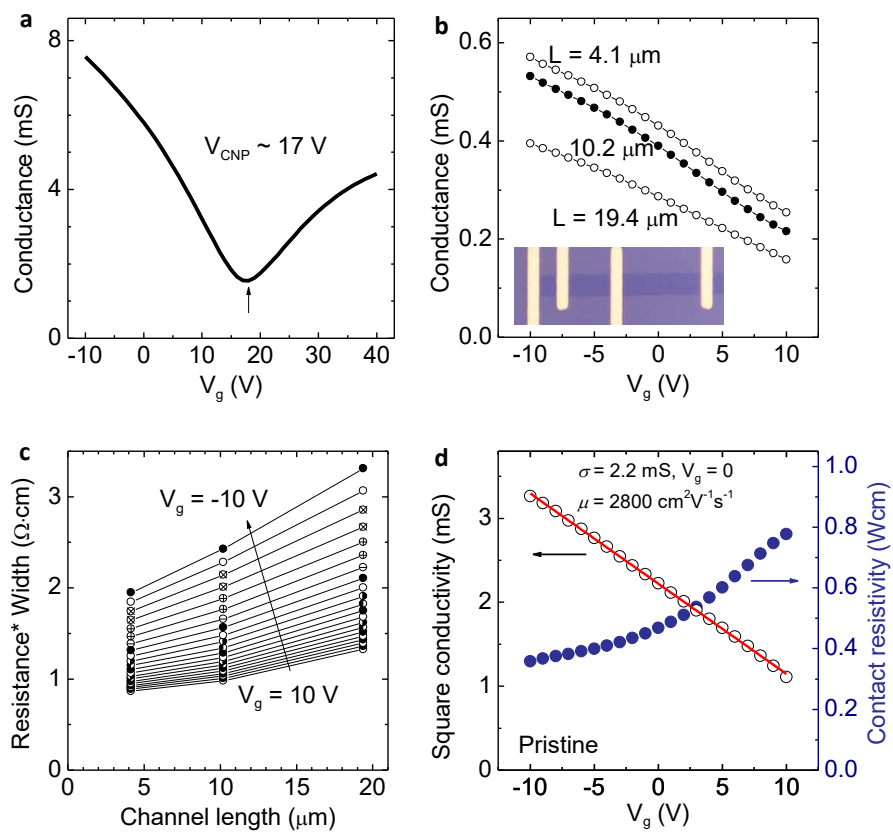

**Figure S5 Extracting intrinsic carrier mobility and contact resistivity of pristine CVD graphene. a**, A typical transfer curve of the pristine CVD graphene in air, where the charge neutrality point ($V_{CNP}$) is normally at $V_g$ around 15-17 V. Here the thickness of $SiO_2$ dielectric is 90 nm and $V_{ds}$ = 0.1 V. **b**, Transfer curves for graphene channels with different channel lengths. Inset: Optical image of the device for TLM measurement. **c**, TLM fit to extract mobility and contact resistance. **d**, Left axis: Intrinsic transfer curve after ruling out the contact effect, which can be used to extract intrinsic mobility. Right axis: Extracted contact resistivity *versus* gate voltage.

The CVD graphene was grown on copper foils and its quality was carefully checked. The intrinsic carrier mobility and contact resistance were estimated by the conventional transfer-length measurement (TLM). In the ambient environment the graphene is p-doped by oxygen and humidity, which shifts the charge neutrality point ($V_{CNP}$) to around $V_g$ = 15-17 V, as shown in Figure S5a. To estimate the hole (the leading carriers passing graphene in this study) mobility of the CVD graphene, we used the data in the $V_g$ range from -10 to 10 V. As shown in the inset of Figure S5b, four electrodes with varied spacing were defined on a well etched graphene ribbon, where three graphene channels were defined with channel lengths (*L*) of 4.1, 10.2 and 19.4 μm, respectively. Figure S5b shows the transfer curves of the three channels where the gate voltage ($V_g$) is changed from -10 to +10 V. All channels exhibit p-doping behavior due to the oxygen doping effect in air. Figure S5c shows the TLM plots at different $V_g$ values at a step of 1V. The intercept and the slope of the linear fits of the data reflect the information of contact resistance and intrinsic channel conductivity (Figure S5d). We extracted an intrinsic carrier mobility ($\mu$) of 2800 cm$^2$V$^{-1}$s$^{-1}$ and a zero-bias square conductivity ($\sigma$) of 2.2 mS, suggesting high quality of the CVD graphene used. The extracted contact resistivity of Cr/Au electrode to graphene is also plotted to the right Y axis in Figure S5d. It increases from 0.35 to 0.8 Ω·cm when $V_g$ changes from -10 to 10V, which is about 10 times higher than that of the Pd/Au contacts (0.02-0.04 Ω·cm).[1] The less perfect contact is likely responsible for the large series resistance of the devices (discussed below).

## 4. Effect of TFSA doping on CVD graphene

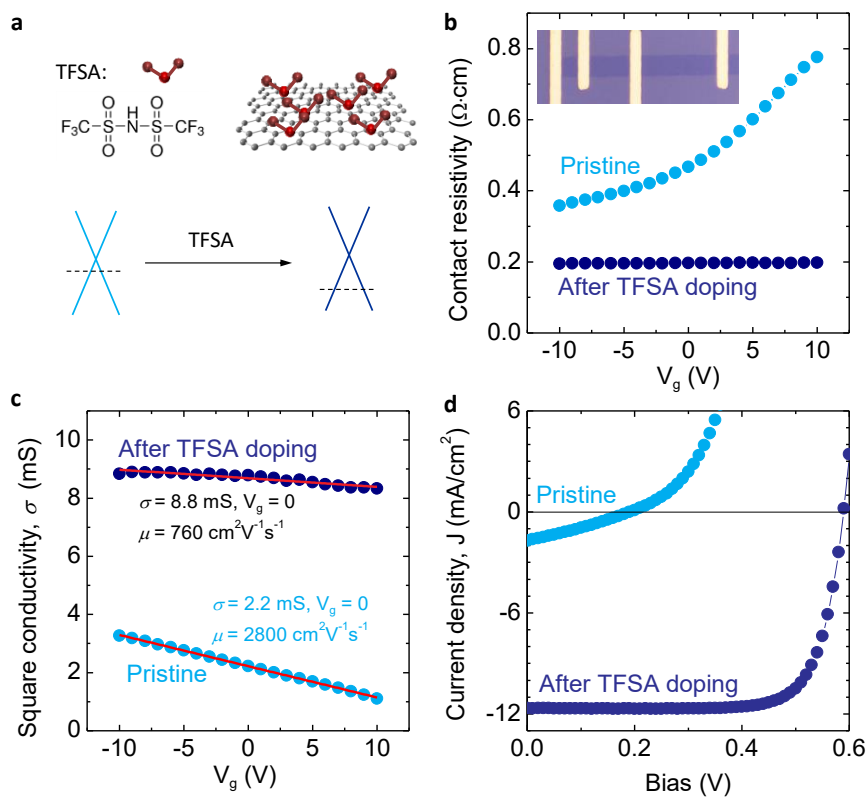

**Figure S6 Effect of TFSA doping on electronic performance of CVD graphene and related photodiodes. a**, Diagram of TFSA doping and Fermi level engineering of graphene. **b**, Variation of contact resistivity by doping. **c**, Change of channel conductivity and carrier mobility by doping. **d**, Improvement of photoresponse after doping.

Figure S6a show the diagram of engineering the Fermi level of graphene. Since TFSA is a strong electron-drawing molecule, upon spinning coating it on graphene, it is expected to highly dope graphene and downshift the Fermi level *via* charge transfer. Electronically, the TFSA doping has three effects: 1) It reduces the contact resistivity to 0.2 Ω·cm in most $V_g$ regime (Figure S6b); 2) It increases $\sigma$ of graphene from 2.2 to 8.8 mS at $V_g$=0; 3) It degrades from 2800 to 760 $cm^2V^{-1}s^{-1}$, because the TFSA molecules are randomly distributed charges above graphene, which serves as additional scattering centers. Although the third effect is detrimental to photoresponse, the residual mobility remains high enough to support the photocarrier transport. After TFSA doping, the overall device performance is much improved by the first two favorable effects. A large increase on both short-circuit current ($J_{sc}$) and open-circuit voltage ($V_{oc}$) is observed (Figure S6d).

## 5. Effect of HNO₃ doping on mechanically exfoliated graphene

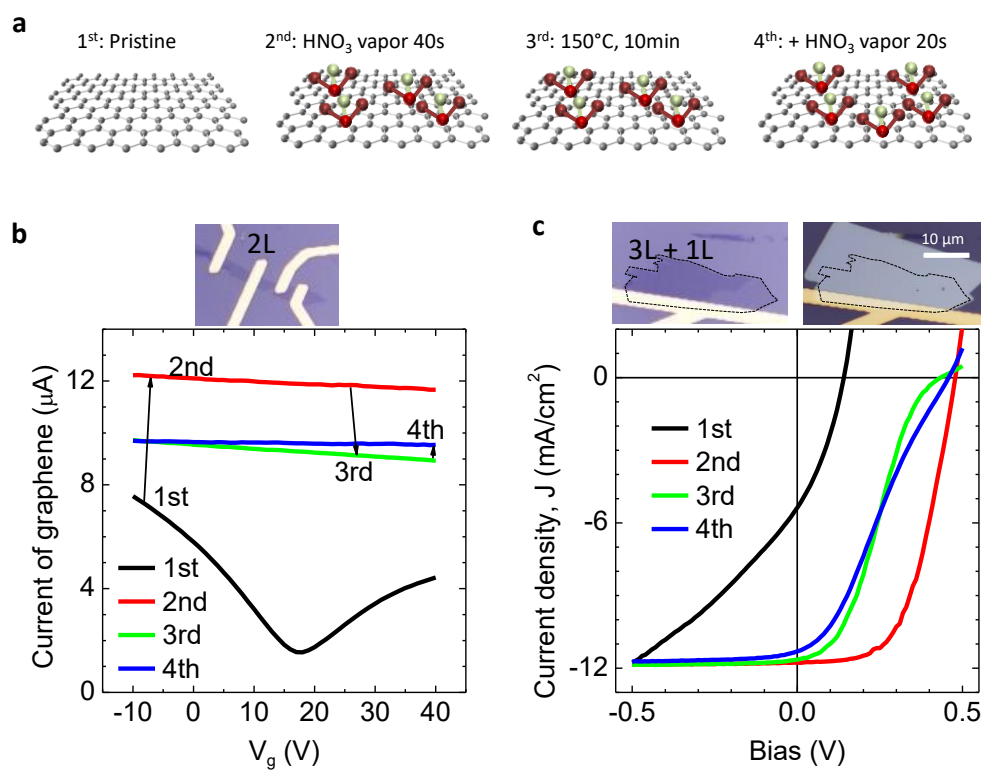

**Figure S7 Effect of HNO₃ doping on electronic performance of mechanically exfoliated graphene and related photodiodes. a**, Diagram of different doping levels. **b**, Transfer curves at different doping levels. **c**, Variation of photoresponse curves at different doping levels.

Figure S7 show sequential experiment results of HNO₃ doping on an individual device at four stages: 1) Pristine; 2) HNO₃ vapor exposure by 40 s; 3) Heated at 150 °C by 10min; 4) Re-exposure of HNO₃ vapor by 40 s. The schematic diagrams are shown in Panel a. Panel b shows the transfer curve at the four conditions, exhibiting quite similar trends with TFSA doping, that is, $\sigma$ increases and $\mu$ decreases upon doping (Stage 1 to 2), reflecting the effect of charge transfer and effective doping. In the stage 3 of heating (150 °C, 10min), $\sigma$ decreases and $\mu$ is unchanged, which means degradation of contact. This is consistent with the S-shape photocurrent curve (green line, panel c). In the stage 4 of re-exposure of HNO₃ vapor, no large improvement in $\sigma$ but further $\mu$ degradation are seen, indicating the saturation in doping. Further exposure to HNO₃ vapor will only increase the defect density in graphene and there is no recovery of the photocurrent curve (blue line, panel c) because HNO₃ can produce carbon vacancies by reaction with graphene.

## 6. Refractive indices of candidate antireflection capping layers

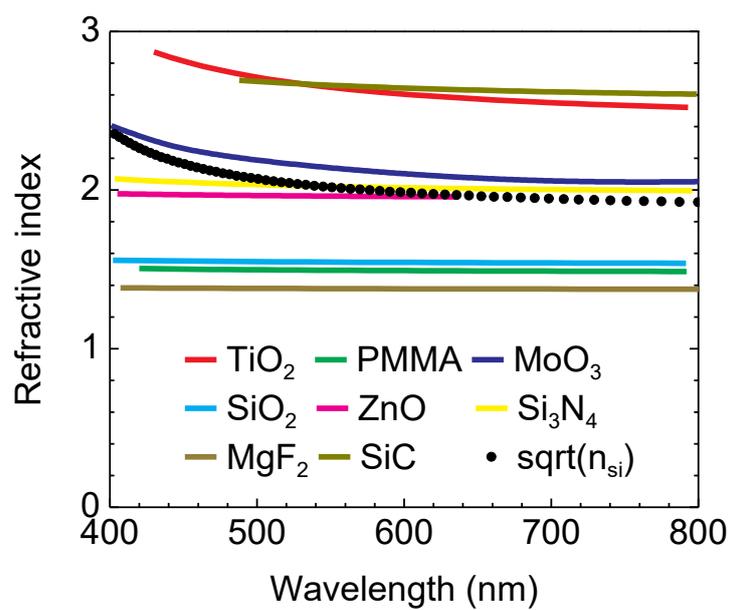

**Figure S8** Refractive index values for several candidate antireflection capping layers and the ideal values ($n_{AR} = \sqrt{n_{air} n_{si}}$).

## 7. Design of MoO$_3$ antireflection layer

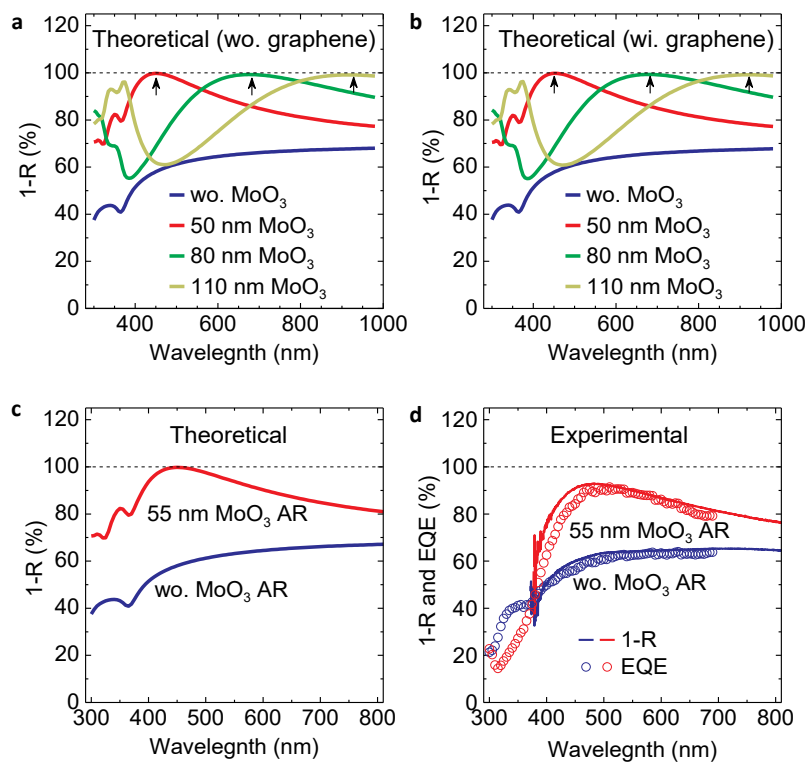

**Figure S9 Design and application of MoO$_3$ antireflection layer. a**, Calculated absorption (1-R) for different AR thicknesses without placing graphene. **b**, Corresponding results with placing graphene. No noticeable difference can be seen between **a** and **b**. **c**, Calculated absorption (1-R) curves for zero and 55 nm MoO$_3$ AR. **d**, Corresponding experimental data on absorption (1-R) and EQE.

The application of antireflection (AR) layers is a common technique to increase the device photoresponse. Theoretical calculation was used for rational design of the thickness of the MoO$_3$ AR layers.

We first identified that the ultrathin layers (such as graphene and TFSA) have negligible effect on the optical reflection. Figure S9 a-b compare the calculated absorption (1-R) curves with and without graphene in the layered devices. As can be seen, the 1-R curves are nearly identical at each MoO$_3$ thickness. Hence, for simplification we neglected the ultrathin layers of graphene and TFSA in later calculations. In Figure S9b, one can find that a MoO$_3$ layer with thickness between 50-80 nm can enhance the 1-R values in most visible range. In experiment, we used 55 nm MoO$_3$.

Figure S9c plots the calculated 1-R curves for devices with bare and 55 nm MoO$_3$

AR layers. Evidently, upon depositing the MoO$_3$ AR layer, the 1-R values are enhanced in the whole visible regime. In particular, the 1-R can be enhanced from 60% to 100% at 450 nm due to the destructive interference in reflection. We also determined the real 1-R curve with reflection spectroscope from 400 to 800 nm (blue and red lines, Figure S9d) and calculated the external quantum efficiency (EQE, blue and red circles, Figure S9d) from 300 to 700 nm. Both values (1-R and EQE) are quite close at each wavelength, implying a nearly unity internal quantum efficiency (IQE = EQE/(1-R)) and a high interface quality of our devices fabricated with the nano-subsidence technique. Overall, the experimental curves shown in panel d agree reasonably with the calculation shown in panel c.

## 8. Effect of barrier height ($\Phi_B$) on $V_{oc}$

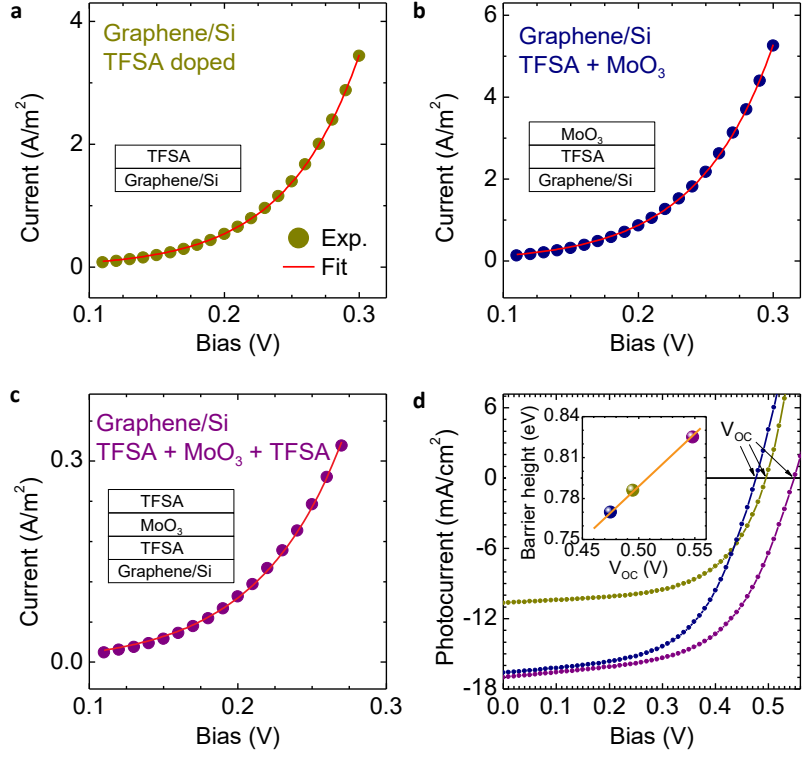

**Figure S10 Dependence of open-circuit voltage $V_{oc}$ on barrier height $\Phi_B$. a-c**, Fits of barrier height by thermionic-emission equation at varied doping levels. Insets: Diagram of doping schemes. **d**, Photoresponse curves at different doping levels where $V_{oc}$ changes accordingly. Inset: Plot of $\Phi_B$ and $V_{oc}$ at different doping levels.

We also investigated the relation of barrier height ($\Phi_B$) and $V_{oc}$. $\Phi_B$ was tuned by changing the work function of graphene *via* doping levels. Three doping levels were achieved by consequently applying TFSA, MoO$_3$, and a second TFSA; the doping schemes were shown in the insets of Figure S10 a-c, respectively. The values of $\Phi_B$ were estimated by fitting the *I-V* curves with the thermionic-emission equation[2]

$$J_{\text{dark}} = A^{**}T^2 \exp\left(-\frac{\Phi_B}{\beta}\right)\left[\exp\left(-\frac{V}{n\beta}\right) - 1\right]$$

where $J_{\text{dark}}$ is the dark current and $A^{**}$=252.4 A·cm$^{-2}$·K$^{-2}$ is the Richardson constant along the <100> direction of silicon. The fits are represented by the red lines in Figure S10 a-c, which agree well with the experimental data. The extracted $\Phi_B$ are 0.786, 0.770, and 0.825 eV for the above doping levels, corresponding to $V_{oc}$ of 0.495, 0.475, and 0.548 V, respectively (Figure S10d). It is found that $V_{oc}$ exhibit a linear dependence on $\Phi_B$, as shown in the inset of Figure S10d.

## 9. Fit of series resistance

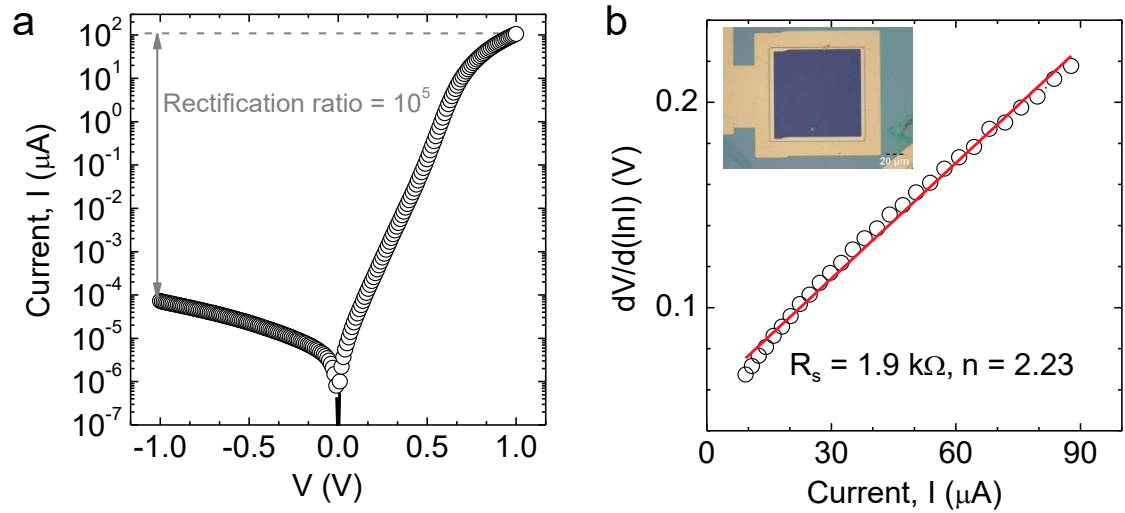

**Figure S11 Fit of series resistance with the transformed thermionic-emission equation. a,** Semi-logarithmic plot of the I-V curves of a TFSA doped graphene/silicon junction under dark condition, where a high rectification ratio of $10^5$ is observed. **b,** Fitting the I-V curve to extract the junction parameters: series resistance $R_s$ and ideality factor $n$. Inset: Optical image of the corresponding device before $SiO_2$ etching.

The series resistance ($R_s$) of the TFSA doped device was fitted with the transformed thermionic-emission equation[3,4]

$$\frac{dV}{d(\ln I)} = R_s I + \frac{n}{\beta}$$

where $n$ is the ideality factor and $\beta = e/kT$ is the inverse thermal voltage. Figure S11 show the experimental data and related fit. The values of $R_s$ and $n$ are estimated to be 1.9 kΩ and 2.23, respectively. Given the dimension of the device (100×100 μm², inset of Figure S11b), the normalized $R_s$ is 0.19 Ω/cm², consistent with previous reports.

## 10. Performance as photovoltaic devices

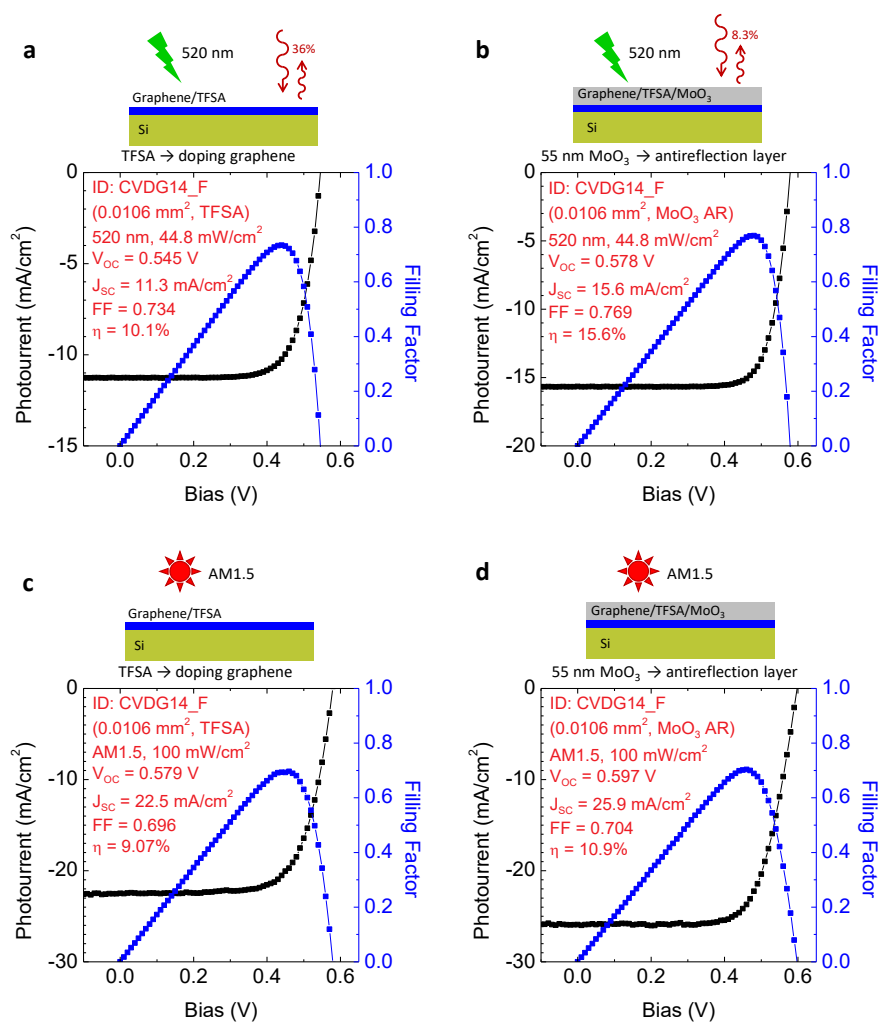

**Figure S12 Comparison of photovoltaic parameters at different device configuration and illumination conditions. a**, No AR layer under monochromatic illumination. **b**, Optimized AR layer under monochromatic illumination. **c**, No AR layer under AM1.5 illumination. **d**, Optimized AR layer under AM1.5 illumination.

Figure S12 show the complete device performance as photovoltaic cells, characterized under different device configuration (with or without $MoO_3$ AR layer) and illumination conditions (monochromatic illumination/ reduced power and standard AM1.5 condition). The detailed device parameters are summarized below.

| Device No. | $MoO_3$ (nm) | Illumination | $V_{oc}$ | $J_{sc}$ | Filling factor, FF | Photoconversion efficiency, $\eta$ |
|---|---|---|---|---|---|---|
| 1 | 0 | 520 nm, 44.8 mW/cm² | 0.545 | 11.3 | 0.734 | 10.1% |
| 2 | 55 |  | 0.578 | 15.6 | 0.769 | 15.6% |
| 3 | 0 | Standard AM1.5 | 0.579 | 22.5 | 0.696 | 9.07% |
| 4 | 55 |  | 0.597 | 25.9 | 0.704 | 10.9% |

## 11. Output power and linearity of monochromatic light source

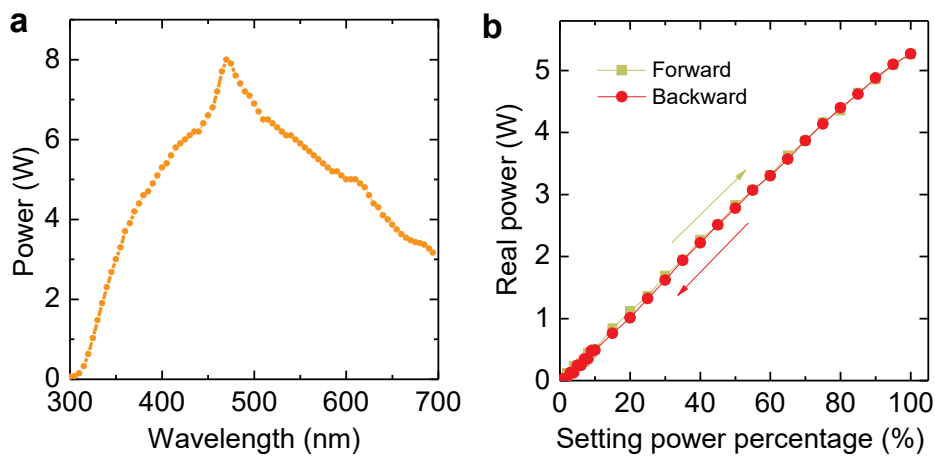

**Figure S13 Output power and linearity of monochromatic light source employed in experiment. a**, Output power as a function of light wavelength where the power peaks at 470 nm. **b**, Test of power linearity of the control system.